%% file: main.tex
\documentclass[twocolumn]{autart}

\usepackage{amsmath}
\usepackage{amssymb} 

\usepackage{custom_SCL}
\usepackage{url}

\begin{document}

\begin{frontmatter}
    \title{
    On Boundedness of Quadratic Dynamics with Energy-Preserving Nonlinearity
    }

    \author[UMich]{Shih-Chi Liao}\ead{shihchil@umich.edu},    
    \author[UMN]{Maziar S. Hemati}\ead{mhemati@umn.edu},  
    \author[UMich]{Peter Seiler}\ead{pjseiler@umich.edu}

    \address[UMich]{Electrical and Computer Engineering, University of Michigan, Ann Arbor, Michigan, USA}  
    \address[UMN]{Aerospace Engineering and Mechanics, University of Minnesota,Minneapolis, Minnesota, USA}             

    \thanks{This material is based upon work supported by the Army Research Office under grant number W911NF-20-1-0156 and the Air Force Office of Scientific Research under grant number FA9550-21-1-0434.}


    \begin{abstract}
    Boundedness is an important property of many physical systems. This includes incompressible fluid flows, which are often modeled by quadratic dynamics with an energy-preserving nonlinearity. For such systems, Schlegel and Noack proposed a sufficient condition for boundedness utilizing quadratic Lyapunov functions. They also propose a necessary condition for boundedness aiming to provide a more complete characterization of boundedness in this class of models.
    The sufficient condition is based on Lyapunov theory and is true. 
    Our paper focuses on this necessary condition. We use an independent proof to show that the condition is true for two dimensional systems. However, we provide a three dimensional counterexample to illustrate that the necessary condition fails to hold in higher dimensions.
    Our results highlight a theoretical gap in boundedness analysis and suggest future directions to address the conservatism.
    \end{abstract}

\end{frontmatter}


\input{Content_Counter/Introduction}

\input{Content_Counter/Preliminaries.tex}

\input{Content_Counter/TwoStates.tex}
\input{Content_Counter/Counterexample.tex}
\input{Content_Counter/Conclusions.tex}
\section*{ACKNOWLEDGMENT}
The authors would like to thank Michael Schlegel and Bernd R. Noack for useful comments and discussions for this work.

\section*{Declaration of Generative AI and AI-assisted technologies in the writing process} \noindent 
Statement: During the preparation of this work the author(s) used GitHub Copilot and Google Gemini in order to edit the article. After using this tool/service, the author(s) reviewed and edited the content as needed and take(s) full responsibility for the content of the publication.

\input{Content_Counter/Appendix.tex}
\bibliographystyle{plain}
\bibliography{Reference.bib}


\end{document}

%% file: Content_Counter/Introduction.tex
\section{Introduction}


Quadratic systems are often used to model complex nonlinear behaviors and serve as higher-order approximations than linearization~\cite{noack2011reduced}.
In some cases, the quadratic terms are known to be energy-preserving (also termed \emph{lossless} in the literature). For example, quadratic systems with energy-preserving nonlinearities arise in reduced-order models of incompressible fluid dynamics, and can exhibit complex behaviors such as chaos and limit cycles~\cite{holmes2012turbulence,noack2011reduced,rowley2017model}.
These models facilitate rapid simulation and analysis, making them valuable for tasks like aerodynamic stability assessment without expensive computations~\cite{taira2020modal}.

The long-term stability of quadratic systems with energy-preserving nonlinearity is crucial for reliable simulations. Boundedness ensures that system behavior remain finite, enabling long-term predictions. However, analyzing boundedness is challenging due to the model nonlinearity. For this class of models, Schlegel and Noack~\cite{schlegel2015long} introduced a sufficient condition for boundedness using quadratic Lyapunov functions and trapping regions. This criterion has been widely adopted in stability analysis~\cite{goyal2025guaranteed,peng2024localstabilityguaranteesdatadriven,liao2025trcvx}, system identification~\cite{Kaptanoglu2021,heide2025datadrivennonlinearaerodynamicsmodels}, and control~\cite{duff2024stability}.

Despite these advances, the necessary condition for boundedness proposed in~\cite{schlegel2015long} remains less understood. Establishing necessary conditions for nonlinear systems is inherently difficult in general~\cite{khalil2002nonlinear}. This motivates a detailed investigation using low-dimensional systems, where analysis is more tractable and insights can be gained. 

The main contributions of this paper are twofold.
First, we verify that the necessary condition for boundedness, proposed by Schlegel and Noack~\cite{schlegel2015long}, holds for two-dimensional systems. We provide an independent proof using a canonical form for the system dynamics.
Second, a three-dimensional counterexample to the necessary condition is presented.
This counterexample illustrates scenarios that may limit the general applicability of the theorem using quadratic Lyapunov functions in systems of dimension greater than two. Note that one-dimensional systems are excluded, as energy-preserving quadratic nonlinearities are trivial in this case.

The paper is organized as follows. Section~\ref{sec:prelim} introduces the problem setup and preliminaries. Section~\ref{sec:2D} presents the two-dimensional characterization of boundedness. Section~\ref{sec:3D} discusses the three-dimensional counterexample and proof investigation. Section~\ref{sec:conclusion} summarizes key findings and future directions.

%% file: Content_Counter/Preliminaries.tex
\section{Preliminaries} \label{sec:prelim}

\subsection{Dynamics with Energy-Preserving Quadratic Terms}
The central object of interest is an $n$-dimensional dynamical system with energy-preserving quadratic terms. We define the system as follows:
\begin{align} \label{eq:general_system}
    \frac{d}{dt}x(t) = c + Lx(t) + \phi(x(t)),
\end{align}
where $x(t)\in\Rn{n}$ is the state at time $t$. The model is defined by the vector $c \in \Rn{n}$, the matrix $L \in \Rn{n \times n}$, and the nonlinearity $\phi: \Rn{n} \to \Rn{n}$. We assume the nonlinearity is energy-preserving: $x^\top \phi(x) = 0$ for all $x \in \Rn{n}$. The energy-preserving assumption is often used to model incompressible fluid flows~\cite{schlegel2015long}, where the kinetic energy is conserved under nonlinear interactions. The nonlinearity $\phi$ can be expressed as:
\begin{align}
    \phi(x) := \begin{bmatrix}
        x^\top Q^{(1)} x \\
        \vdots \\
        x^\top Q^{(n)} x
    \end{bmatrix},
\end{align}
where $Q^{(i)} \in \mathbb{R}^{n \times n}$ is a symmetric matrix for $i=1,\ldots,n$. The energy-preserving property, $x^\top \phi(x) = 0$, implies that the matrices $Q^{(i)}$ satisfy the following conditions:
\begin{align} \label{eq:energy_preserving_Qs}
    Q^{(i)}_{jk} + Q^{(j)}_{ik} + Q^{(k)}_{ij} = 0, \quad \forall i,j,k = 1,\ldots,n. 
\end{align}
A detailed derivation of~\eqref{eq:energy_preserving_Qs} can be found in~\cite{schlegel2015long}.

Coordinate translations play an important role in the analysis below. Specifically, define $y:=x-m$ where $m\in \Rn{n}$ is the coordinate shift. The system dynamics under this transformation take the following form~\cite{schlegel2015long}:
\begin{align} \label{eq:shifted_system}
    \frac{d}{dt}y(t) = d(m) + A(m)y(t) + \phi(y(t)),
\end{align}
where $d(m) := c + Lm + \phi(m)$ is the constant term, and the linear part $A(m)$ is given by:
\begin{align} \label{eq:linear_part}
    A(m) := L + 
    2\begin{bmatrix}
        m^\top Q^{(1)} \\ 
        \vdots \\
        m^\top Q^{(n)}
    \end{bmatrix}.
\end{align}
Note that the transformed system in~\eqref{eq:shifted_system} has a quadratic nonlinearity $\phi(y)$ in the same form. Hence, this shift-transformed system is still energy-preserving.

\subsection{Boundedness}
This paper focuses on the boundedness properties for all trajectories $x(t)$ starting from initial condition $x(t_0) = x_0$. Without loss of generality, we assume $t_0=0$ as the system is time-invariant. The boundedness of the system serves as a notion to characterize the long-term behavior of the system. It is defined in Chapter 4.8 of \cite{khalil2002nonlinear} as the following:
\begin{defn}[Boundedness] \label{def:boundedness}
     A system is globally uniformly ultimately bounded if there exists a constant $\beta > 0$ and a function $T: \mathbb{R}^+ \to \mathbb{R}^+$ such that all trajectories $x(t)$ satisfy:
     \begin{align}
        \norm{x(t)}_2 \leq \beta, \quad \forall x_0 \in \Rn{n} \text{ and } t \geq T(\norm{x_0}_2).
     \end{align}
\end{defn}
In other words, a system is globally uniformly ultimately bounded if all trajectories eventually converge to a bounded region of the state space. Definition~\ref{def:boundedness} specifically states that all trajectories should eventually converge to a ball of radius $\beta > 0$. The time to converge $T$ can depend on the norm of the initial condition $x_0$. In this paper, we will simply refer to this property as \emph{boundedness}.

Note that a coordinate shift $m$ does not affect the long-term behavior of the system~\cite{liao2025trcvx}. Hence boundedness of the shifted system~\eqref{eq:shifted_system} is equivalent to boundedness of the original system~\eqref{eq:general_system}. Therefore we use \emph{boundedness} interchangeably for the original and shifted systems.

\subsection{Sufficient Theorem for Boundedness}
Schlegel and Noack~\cite{schlegel2015long} proposed a sufficient condition for boundedness of energy-preserving quadratic dynamics using the notion of a trapping region. The analysis extends prior work from Lorenz~\cite{lorenz1963deterministic}. This subsection summarizes the condition by Schlegel and Noack~\cite{schlegel2015long}. We also present a convex optimization formulation from~\cite{liao2025trcvx} to test feasibility of the condition. We start with the following definition for trapping regions. 

\begin{defn}[Trapping Region]
    A trapping region $D \subseteq \Rn{n}$ is a compact set that is forward invariant under the dynamics~\eqref{eq:general_system}, i.e., if $x(t_0) \in D$, then $x(t) \in D$ for all $t \geq t_0$. A trapping region is termed globally monotonically attracting if an energy function $K(x):=(x-m)^\top (x-m)$ for some $m\in\Rn{n}$ is strictly monotonically decreasing along all trajectories starting from an arbitrary state outside $D$.
\end{defn}
This paper focuses on a \emph{globally monotonically attracting} trapping region. Hence, we will refer to it simply as a trapping region.
The existence of a trapping region implies the boundedness of the system, as all trajectories will eventually converge to and stay within the trapping region.

The energy function $K(y) := y^\top y$ with $y = x - m$ plays a key  role in the trapping region analysis condition given below for boundedness. The time derivative of this energy function along trajectories of the shifted system is:
\begin{align} \label{eq:energy_evolution}
    \begin{split}
        \frac{d}{dt} K(y(t)) &= y(t)^\top \left(\frac{d}{dt}y(t)\right) + \left(\frac{d}{dt}y(t)\right)^\top y(t) \\
        &= d(m)^\top y(t)  + y(t)^\top A_s(m) y(t),
    \end{split}
\end{align}
where $A_s(m)$ is the symmetric part of $A(m)$:
\begin{align} \label{eq:shifted_symmetric_linear_part}
\begin{split}    
    A_s(m) &:= \frac{1}{2}(A(m) + A(m)^\top) \\
     & = \frac{1}{2}(L + L^\top) - \sum_{i=1}^{n} m_i Q^{(i)}.
\end{split}
\end{align}
Note that $y(t)^\top \phi(y(t))=0$ holds due to energy-preserving properties of $\phi$. As a result, the time derivative of $K(y(t))$ is not affected by the nonlinearity as shown in~\eqref{eq:energy_evolution}.
Moreover, the derivation of $A_s(m)$ utilizes the energy-preserving properties of the nonlinearity $\phi$ given in \eqref{eq:energy_preserving_Qs}~\cite{schlegel2015long}.

Observe that $\frac{d}{dt}K(y(t))$ is dominated by the term $y(t)^\top A_s(m) y(t)$ for large enough $\norm{y(t)}_2$. Therefore, if there exists a shift vector $m$ such that $A_s(m)$ is negative definite, then $\frac{d}{dt}K(y(t)) < 0$ for sufficiently large $\norm{y(t)}_2$. This implies that all trajectories will eventually enter and stay within a bounded region of the state space, i.e., a trapping region exists. A semidefinite program (SDP) can be used to find such a trapping region (if one exists) by minimizing the largest eigenvalue of the symmetric linear part $A_s(m)$ of the shift-transformed system:
\begin{align} \label{SDP:TR_sufficient}
    a^* = \min_{m \in \Rn{n},\, a \in \Rn{}} a \quad \text{s.t.} \quad A_s(m) \preceq a I.
\end{align}
The next theorem provides a sufficient condition for the existence of a trapping region for dynamics~\eqref{eq:general_system}:
\begin{thm} \label{thm:boundedness_sufficient}
    Consider system~\eqref{eq:general_system} with energy-preserving quadratic dynamics and let $a^*$ denote the optimal value of the SDP~\eqref{SDP:TR_sufficient}.
    If $a^* < 0$, then a trapping region exists for~\eqref{eq:general_system} and the system is bounded.
    If $a^* \geq 0$, no trapping region exists.
\end{thm}

The proof of Theorem~\ref{thm:boundedness_sufficient} can be found in~\cite{liao2025trcvx} along with further characterization of the trapping region if one exists. The proof of Theorem~\ref{thm:boundedness_sufficient} uses a Lyapunov-like argument to show that the energy function $K$ is decreasing along trajectories of the translated system~\eqref{eq:shifted_system} when $\norm{y(t)}_2$ is sufficiently large. The convex optimization~\eqref{SDP:TR_sufficient} provides a computationally efficient condition to analyze the boundedness of lossless quadratic systems.

\subsection{Necessary Theorem for Boundedness}
Schlegel and Noack~\cite{schlegel2015long} also state a necessary condition, under technical assumptions, for quadratic systems with energy-preserving nonlinearity. The necessary condition involves the concept of an effective nonlinearity.  We restate the definition below with a slight modification for clarity.
\begin{defn}[Effective Nonlinearity] \label{def:effective_nonlinearity}
    The nonlinearity in the system~\eqref{eq:general_system} is not effective if there exists a non-trivial linear subspace, $\mathcal{V} \subset \Rn{n}$, such that:
    \begin{enumerate}
        \item $\phi(x)=0$ for all $x \in \mathcal{V}$, and
        \item $c+Lx \in \mathcal{V}$ for all $x \in \mathcal{V}$.
    \end{enumerate}
    The system has an \emph{effective nonlinearity} if no such non-trivial linear subspace exists.
\end{defn}
The first condition is equivalent to the nonlinearity $\phi(x)$ vanishing in the subspace. The second condition is equivalent to the affine and linear dynamics $c+Lx$ being invariant in the subspace, i.e., the remaining dynamics do not leave the subspace~\cite[Theorem 2.4]{trentelman2012control}. Thus, if the nonlinearity is not effective, then $x(0) \in \mathcal{V}$ implies the $x(t) \in \mathcal{V}$ for all $t \geq 0$. Moreover, the trajectory follows the affine and linear dynamics on this subspace. An effective nonlinearity precludes this special case and ensures that the nonlinearity has some effect on all trajectories of the system.



Theorem 2 in Schlegel and Noack~\cite{schlegel2015long} provides the necessary condition for boundedness using this effective nonlinearity concept. This result is restated in the next theorem.
\begin{thm} \label{thm:necessity} 
    Consider a system~\eqref{eq:general_system} with an effective nonlinearity. If the system is bounded, as given in Definition~\ref{def:boundedness}, then there exists a shift vector $m$ such that $A_s(m) \preceq 0$.
\end{thm}

The proposed necessary theorem aims to complement the sufficient condition in Theorem~\ref{thm:boundedness_sufficient} and provide a more comprehensive understanding of the relationship between system dynamics and boundedness. 

\subsection{Problem Statement and Overview}

Theorem~\ref{thm:boundedness_sufficient} is based on a Lyapunov-like argument on the translated system. This theorem is true and is not the focus of our paper. 
Instead, we focus on Theorem~\ref{thm:necessity} and show that this result is false in certain cases. Specifically, we show in Section~\ref{sec:2D} that Theorem~\ref{thm:necessity} is true for two-dimensional systems. We use an independent proof unrelated to the one used in~\cite{schlegel2015long}. Next, we present a three-dimensional system in Section~\ref{sec:3D} and show that this system is a counterexample to Theorem~\ref{thm:necessity}. We can always augment this system with additional dynamics to generate counterexamples for higher dimensions. This implies that Theorem~\ref{thm:necessity} is false, in general, for $n\ge 3$. Note that systems with one state are not considered as any energy preserving quadratic nonlinearity is trivial, i.e., $\phi(x) = 0$ for all $x\in \Rn{}$.

%% file: Content_Counter/TwoStates.tex
\section{Analysis of Two-dimensional Systems} \label{sec:2D}



In this section, we examine the simplest non-trivial case of energy-preserving quadratic systems: systems with two states. First, a canonical form is developed that can be considered without loss of generality. Theorem~\ref{thm:necessity} is then examined using this canonical form and verified to be true for two-dimensional systems.

Consider the quadratic system \eqref{eq:general_system} with two states, $x\in \Rn{2}$.  The energy-preserving condition \eqref{eq:energy_preserving_Qs} implies that the two matrices in $\phi$ can be parameterized as:
\begin{align}
    Q^{(1)} = \begin{bmatrix}
        0 & 0.5q_1 \\
        0.5q_1 & q_2
    \end{bmatrix}, \quad 
    Q^{(2)} = \begin{bmatrix}
        -q_1 & -0.5q_2 \\
        -0.5q_2 & 0
    \end{bmatrix},
\end{align}
where $q_1$, $q_2 \in \Rn{}$. Therefore, $\phi$ can be expressed as:
\begin{align} \label{eq:phi2D}
    \phi(x) = \bmtx x^\top Q^{(1)} x \\ x^\top Q^{(2)} x \emtx 
                = (q^\top x) \bmtx 0 & 1 \\ -1 & 0 \emtx x,
\end{align}
where $q:= \bsmtx q_1 \\ q_2 \esmtx \in \Rn{2}$. We can verify the lossless condition $x^\top \phi(x) =0$ due to the skew-symmetric matrix in $\phi$.  Equation~\ref{eq:phi2D} gives the general form for the nonlinearity in this class of systems.

\subsection{Canonical Form}

To simplify further we consider the effect of a coordinate rotation:
\begin{align}
    \hat{x} := R x \, \mbox{ where } R^\top R = I_2.
\end{align}
Differentiating $\hat{x}(t)$ along trajectories of the original system~\eqref{eq:general_system} gives:
\begin{align}
    \frac{d}{dt} \hat{x}(t) &= R \left[ c + L x(t) + \phi(x(t)) \right].
\end{align}
The dynamics in rotated coordinates are obtained by substituting $x(t)=R^\top \hat{x}(t)$ into the right side and using the form of $\phi$ in \eqref{eq:phi2D}. This yields:
\begin{align} \label{eq:rotated2D}
    \frac{d}{dt} \hat{x}(t) 
    & = \hat{c} + \hat{L} \hat{x}(t) + (\hat{q}^\top \hat{x}(t)) \bmtx 0 & 1 \\ -1 & 0 \emtx \hat{x}(t),
\end{align}
where $\hat{c} := R c$, $\hat{L}:=R LR^\top$, and $\hat{q}:= R q$.  The definition of $\hat{q}$ follows because
\begin{align}
    R  \bmtx 0 & 1 \\ -1 & 0 \emtx R^\top =  \bmtx 0 & 1 \\ -1 & 0 \emtx.
\end{align}
The coordinate change $\hat{x}(t)=Rx(t)$ only rotates trajectories. Hence, the quadratic system~\eqref{eq:general_system} with two states is bounded if and only if the rotated system in \eqref{eq:rotated2D} is bounded. The next lemma states that the trapping region condition is also invariant to rotations for two-dimensional systems.

\begin{lem} \label{lem:2D_rotation_invariance}
Consider a quadratic system~\eqref{eq:general_system} with two states. Let $a^*$ and $\hat{a}^*$ denote the optimal costs of the trapping region SDP~\eqref{SDP:TR_sufficient} using the data from the original~\eqref{eq:general_system} and rotated~\eqref{eq:rotated2D} systems, respectively. Then $a^*=\hat{a}^*$.
\end{lem}

The proof of this lemma is given in Appendix~\ref{sec:appendix_2DLemmaProof}. In summary, boundedness and the trapping region condition~\eqref{SDP:TR_sufficient} are both invariant to rotations when the system is two-dimensional. Thus, we can use a rotation to reduce the system to a canonical form that simplifies the analysis. Specifically, consider a non-trivial quadratic nonlinearity defined by $q\in\Rn{2}$ with $q_0:=\norm{q}_2\neq 0$. Then, we can define a rotation $R$ such that $\hat{q} = R q = q_0 \bsmtx 1 \\ 0 \esmtx$. The canonical form is thus:
\begin{align*} 
  \frac{d}{dt} \hat{x}(t) = \hat{c} + \hat{L} \hat{x}(t) + q_0 \bmtx \hat{x}_1(t) \hat{x}_2(t) \\ -\hat{x}_1^2(t) \emtx.
\end{align*}
The corresponding matrices in the quadratic nonlinearity are:
\begin{align}
    \hat{Q}^{(1)} = q_0 \begin{bmatrix}
        0 & 0.5 \\
        0.5 & 0
    \end{bmatrix}, \quad 
    \hat{Q}^{(2)} = q_0 \begin{bmatrix}
        -1 & 0 \\
        0 & 0
    \end{bmatrix}.
\end{align}
Every two-dimensional system (with nontrivial nonlinearity) can be rotated to this canonical form with no effect on boundedness or the trapping
region condition. The remainder of this section assumes, without loss of generality, that the system is in this
canonical form, and the hat notations are dropped for simplicity:
\begin{align} \label{eq:2D_system}
    \frac{d}{dt}x(t) = c + Lx(t) + q_0 \begin{bmatrix}
        x_1(t)x_2(t) \\ -x_1^2(t)
    \end{bmatrix}.
\end{align}

\subsection{Proof of Theorem~\ref{thm:necessity} for $n=2$}

To start, we consider the LMI condition $A_s(m) \preceq 0$ that appears in the conclusion of Theorem~\ref{thm:necessity}.
This is equivalent, for the canonical system, to the following LMI being feasible:
\begin{align} \label{eq:2D_LMI}
    A_s(m) = \begin{bmatrix}
        l_{11} + q_0 m_2 & \frac{1}{2}(l_{12}+ l_{21} - q_0 m_1) \\
        \frac{1}{2}(l_{12}+ l_{21} - q_0 m_1) & l_{22}
    \end{bmatrix} \preceq 0,
\end{align}
where $l_{ij}$ and $m_i$ are the elements of $L$ and $m$. The following lemma examines the feasibility of this LMI and its relationship to system parameters.
\begin{lem} \label{lem:2D_LMI_infeasible}
    The LMI~\eqref{eq:2D_LMI} is infeasible for all $m\in\Rn{2}$ if and only if $l_{22} > 0$.
\end{lem}
\begin{pf}
    If $l_{22} > 0$ then the LMI~\eqref{eq:2D_LMI} cannot be satisfied for any $m$ as the lower right entry is positive. Conversely, if $l_{22} \leq 0$, then we can choose $m \in \Rn{2}$ to ensure the LMI~\eqref{eq:2D_LMI} is feasible. Specifically, the choice of $m_1 = \frac{1}{q_0}(l_{12} + l_{21})$ and $m_2 = -\frac{1}{q_0}(l_{11} + \epsilon)$ for some $\epsilon > 0$ yields
    \begin{align}
        A_s(m) = \begin{bmatrix}
            -\epsilon & 0 \\
            0 & l_{22}
        \end{bmatrix} \preceq 0.
    \end{align}
    Hence, the LMI~\eqref{eq:2D_LMI} is infeasible if and only if $l_{22} > 0$.
    $\blacksquare$
\end{pf}

The next lemma examines the dynamics with $l_{22} > 0$ and shows that the system is unbounded in this case.
\begin{lem} \label{lem:2D_unbounded_l22_gt_0}
    If $l_{22} > 0$, then the system~\eqref{eq:2D_system} is unbounded.
\end{lem}
\begin{pf}
    The time-derivative of the second state in the canonical system~\eqref{eq:2D_system} is:
    \begin{align}
        \frac{d}{dt}x_2(t) = c_2 + l_{21} x_1(t) + l_{22} x_2(t) - q_0 x_1^2(t).
    \end{align}
    This can be rewritten by completing the square as:
    \begin{align*}
        \frac{d}{dt}x_2(t) &= -q_0 \left( x_1(t) - \frac{1}{2q_0} l_{21} \right)^2 + \left( c_2 + \frac{1}{4q_0} l_{21}^2 \right) \\ &\qquad + l_{22} x_2(t).
    \end{align*}
    Next define the constant $k:= c_2 + \frac{1}{4q_0} l_{21}^2 + 1$.
    Since $l_{22}>0$ by assumption, it follows that if $x_2(t)<-\frac{k}{l_{22}}$ then the derivative is bounded as follows:
    \begin{align}
        \frac{d}{dt}x_2(t) < -q_0 \left( x_1(t) - \frac{1}{2q_0} l_{21} \right)^2 - 1 \le -1
    \end{align}
    This implies that if $x_2(0)<-\frac{k}{l_{22}}$ then $x_2(t) < x_2(0)-t$. In other words, if $x_2(0)$ is sufficiently small then it will grow unbounded, $x_2(t)\to -\infty$ as $t\to \infty$.
    $\blacksquare$
\end{pf}

The next theorem connects these lemmas:
\begin{thm} \label{thm:necessity_2D}
    If the system~\eqref{eq:2D_system} is bounded, then the LMI~\eqref{eq:2D_LMI} is feasible.
\end{thm}
\begin{pf}
    The proof is by contraposition.
    If the LMI~\eqref{eq:2D_LMI} is infeasible, then $l_{22} > 0$ by Lemma~\ref{lem:2D_LMI_infeasible}. If $l_{22} > 0$, then the system is unbounded by Lemma~\ref{lem:2D_unbounded_l22_gt_0}.
    It follows that infeasibility of the LMI~\eqref{eq:2D_LMI} implies that the system is unbounded. By contraposition, if the system~\eqref{eq:2D_system} is bounded, then the LMI~\eqref{eq:2D_LMI} is feasible.
        $\blacksquare$
\end{pf}

Theorem~\ref{thm:necessity_2D} verifies that the necessary condition in Theorem~\ref{thm:necessity} is true for two-dimensional systems. Our proof is independent of the one given in~\cite{schlegel2015long} and relies on the canonical form for $n=2$.



%% file: Content_Counter/Counterexample.tex
\section{A Three-dimensional Counterexample} 
\label{sec:3D}

Consider the following three-state system with lossless quadratic dynamics:
\begin{align} \label{eq:counterexample}
    \frac{d}{dt}\begin{bmatrix}
        x_1(t) \\
        x_2(t) \\
        x_3(t)
    \end{bmatrix} &= Lx(t) + \phi(x(t)), 
\end{align}
where
\begin{align}
\label{eq:Lphi3}
    L &:= \begin{bmatrix}
        -2 & 1 & 0 \\
        -1 & 0.5 & 3 \\
        0 & -3 & -3
    \end{bmatrix}, &
    \phi(x) &:= \begin{bmatrix}
        x_2x_3 \\
        -x_1x_3 \\
        0 
    \end{bmatrix}.
\end{align}
The nonlinearity is energy-preserving as it satisfies:
\begin{align}
    x^\top \phi(x) = 0, \quad \forall x \in \mathbb{R}^3.
\end{align}
%
\begin{figure*}[t]
    \centering
    \includegraphics[width=1.07\columnwidth]{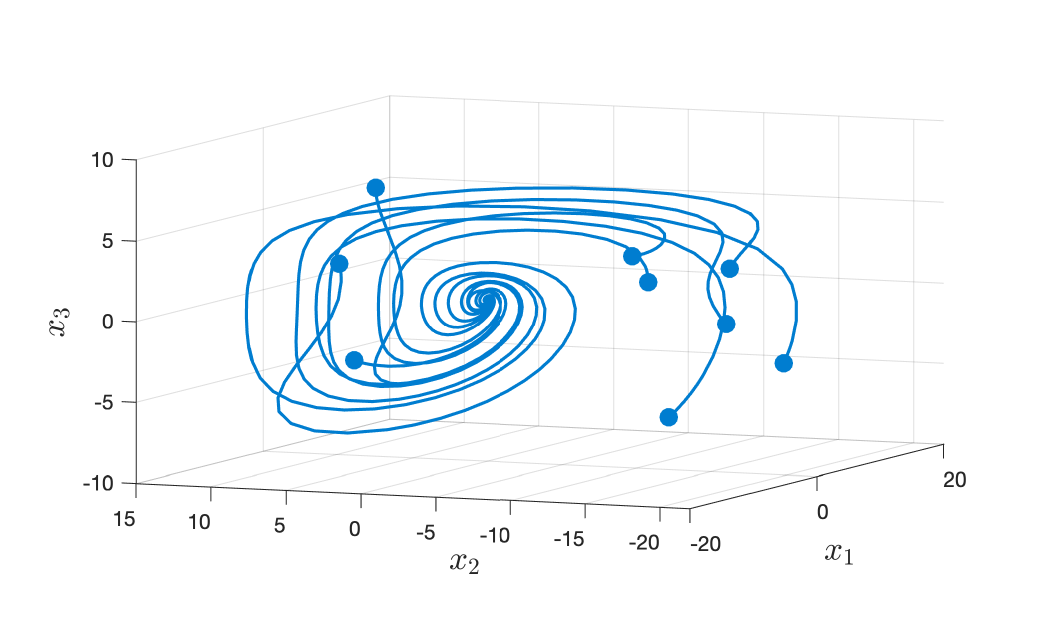}
    \includegraphics[width=0.93\columnwidth]{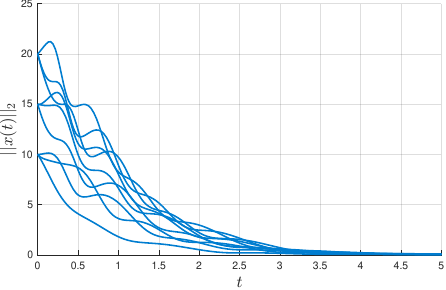}
    \caption{Simulations of the three-dimensional system~\eqref{eq:counterexample}. The left plot shows the trajectories of the states $x_1(t)$, $x_2(t)$, and $x_3(t)$ starting from random initial conditions. All trajectories converge to the origin, illustrating global asymptotic stability and long-term boundedness. The right plot shows the Euclidean norm of the state vector, $\|x(t)\|_2$, over time. The norm converges to zero for all trajectories.
    }
    \label{media:3D_Counterexample_Simulation}
\end{figure*}

Figure~\ref{media:3D_Counterexample_Simulation} shows several simulations of the system~\eqref{eq:counterexample} starting from random initial conditions. Both three-dimensional state-space trajectories and the Euclidean norm of the state vector over time are shown. All simulated trajectories converge to the origin. We will formally show that the origin is a globally asymptotically stable equilibrium point in Lemma~\ref{lemma:Counterexample_LongTermBoundedness}. This illustrates the boundedness of the system.

In the remainder of this section, we show that this system provides a counterexample to Theorem~\ref{thm:necessity}. We also discuss a potential issue in the proof provided in~\cite{schlegel2015long}.

\subsection{Verification of Counterexample}
In this subsection, we show that the system~\eqref{eq:counterexample} is a counterexample to the Theorem~\ref{thm:necessity}. We first show that the origin $x=0$ is globally asymptotically stable and hence the system is bounded. Next, we show that the nonlinearity is effective. Finally, we show that $A_s(m)$ has a positive eigenvalue for any coordinate shift $m \in \Rn{3}$. We combine these steps to verify that the system~\eqref{eq:counterexample} is a counterexample to the Theorem~\ref{thm:necessity}.

The next lemma verifies the stability and boundedness of the system~\eqref{eq:counterexample}.

\begin{lem} \label{lemma:Counterexample_LongTermBoundedness}
    The origin $x=0$ is a globally exponentially stable equilibrium point of the system~\eqref{eq:counterexample}, and the system is bounded.
\end{lem}

\begin{pf}
    Consider the following quartic, Lyapunov function candidate:
    \begin{align}
        V(x) := z(x)^\top M_v z(x), \quad z(x) := \begin{bmatrix}
            x_1 \\ x_2 \\ x_3 \\ x_3^2
        \end{bmatrix},
    \end{align}
    where $M_v$ is defined as:
    \begin{align}
        M_v := 
        \begin{bmatrix}
            136 & 0 & 0 & 6 \\
            0 & 100 & 25 & 0 \\
            0 & 25 & 70 & 0 \\
            6 & 0 & 0 & 1
        \end{bmatrix}.
    \end{align}
    The matrix $M_v$ is positive definite since its eigenvalues are $\{0.7339, 55.85, 114.2, 136.3\}$. It follows that $V$ is a positive definite and decrescent function. It can be verified, with calculus, that the time derivative of $V(x(t))$ along trajectories of the system is given by:
    \begin{align}
        \label{eq:Vdot3D}
        \frac{d}{dt}V(x(t)) = - z(x(t))^\top M_d z(x(t)),
    \end{align}
    where $M_d$ is defined as:
    \begin{align}
        M_d := \begin{bmatrix}
            544 & -36 & 25 & 73 \\
            -36 & 50 & -27.5 & -6 \\
            25 & -27.5 & 270 & 0 \\
            73 & -6 & 0 & 12
        \end{bmatrix}.
    \end{align}
    To show the origin is exponentially stable, rewrite Equation~\ref{eq:Vdot3D} as follows
    \begin{align}
        \label{eq:Vdotequality}
        \frac{d}{dt}V(x(t)) = -\alpha V(x(t)) + z(x(t))^\top N z(x(t)),
    \end{align}
    where $\alpha = 0.1$ and $N := -M_d + \alpha M_v$ is given by:
    \begin{align}
        N = \begin{bmatrix}
            -530.4 & 36 & -25 & -72.4 \\
            36 & -40 & 30 & 6 \\
            -25 & 30 & -263 & 0 \\
            -72.4 & 6 & 0 & -11.9
        \end{bmatrix}.
    \end{align}
    The eigenvalues of $N$ are given by $\big\{-1.874, -33.95, \allowbreak -2263.9, -545.5\big\}$. Hence, $N$ is negative semi-definite and Equation~\eqref{eq:Vdotequality} implies the following inequality:
    \begin{align}
    \frac{d}{dt}V(x(t)) \leq -\alpha V(x(t))
    \end{align}
    By the Lyapunov stability theorem~\cite[Theorem 4.10]{khalil2002nonlinear}, the origin of the system is globally exponentially stable. All trajectories converge to the origin, and therefore the system is bounded.
    $\blacksquare$
\end{pf}

The next lemma shows that the system~\eqref{eq:counterexample} has an effective nonlinearity (Definition~\ref{def:effective_nonlinearity}),

\begin{lem} \label{lemma:Counterexample_EffectiveNonlinearity}
    The system~\eqref{eq:counterexample} has an effective nonlinearity.
\end{lem}

\begin{pf}
    Recall the nonlinearity $\phi$ is not effective if a non-trivial linear subspace, $\mathcal{V} \subset \Rn{3}$ exists satisfying two conditions:
    \begin{enumerate}
        \item $\phi(x)=0$ for all $x \in \mathcal{V}$, and
        \item $c+Lx \in \mathcal{V}$ for all $x \in \mathcal{V}$.
    \end{enumerate}
    This proof verifies that no such linear subspace exists for the system~\eqref{eq:counterexample}.

    First, the nonlinearity $\phi$ defined in \eqref{eq:Lphi3}
    vanishes in the following four candidate linear subspaces:
    \begin{align}
        \mathcal{V}_i &= \left\{x_i e_i \,\, | \,\, x \in \mathbb{R}^3\right\}, \quad i={1,2,3}, \\
        \mathcal{V}_4 &= \left\{[x_1, x_2, 0]^\top 
        \,\, | \,\,  x_1, x_2 \in \mathbb{R}\right\},
    \end{align}
    where $e_i$ is the $i$-th standard basis vector in $\mathbb{R}^3$. Note that $\mathcal{V}_1 \subset \mathcal{V}_4$ and $\mathcal{V}_2 \subset \mathcal{V}_4$. These four subspaces are the only candidates that satisfy the first condition, as $\phi(x) = 0$ if and only if $x_3 = 0$ or $x_1 = x_2 = 0$.

    Next, the candidate subspaces are checked against the second condition. The system \eqref{eq:counterexample} has $c=0$ and hence this condtion simplifies to:
    $Lx \in \mathcal{V}$ for all $x \in \mathcal{V}$.  

    Consider the first subspace $\mathcal{V}_1$. The first basis vector $e_1 := \bmtx 1 & 0 & 0 \emtx^\top$ is in $\mathcal{V}_1$. However, $Le_1 = \bmtx -2 & -1 & 0 \emtx^\top$ is not in $\mathcal{V}_1$. Therefore, $\mathcal{V}_1$ fails to satisfy the second condition. Similarly, it can be shown that $\mathcal{V}_2$ and $\mathcal{V}_3$ do not satisfy the second condition. Next, note that the second basis vector $e_2 := \bmtx 0 & 1 & 0 \emtx^\top$ is also in $\mathcal{V}_4$. However, $Le_2 = \bmtx 1 & -0.5 & -3 \emtx^\top$ is not in $\mathcal{V}_4$. Therefore, $\mathcal{V}_4$ also fails to satisfy the second condition.
    
    Since no linear subspace satisfies both conditions, the nonlinearity is effective.        
    $\blacksquare$
\end{pf}


Finally, we show that $A_s(m)$ has a positive eigenvalue for any $m \in \Rn{3}$.

\begin{lem} \label{lemma:Counterexample_Eigenvalues}
    The matrix $A_s(m)$ has a positive eigenvalue for any $m \in \Rn{3}$.
\end{lem}
\begin{pf}
    The symmetric linear part of the shift-transformed system~\eqref{eq:counterexample} is given by:
     \begin{align}
        A_s(m) &= \frac{1}{2}(L + L^\top) - \sum_{i=1}^{3} m_i Q^i \\
        &= \begin{bmatrix}
            -2 & 0 & 0.5m_2 \\
            0 & 0.5 & -0.5m_1 \\
            0.5m_2 & -0.5m_1 & -3
        \end{bmatrix}.
    \end{align}
    If $x = [0, 1, 0]^\top$ then $x^\top A_s(m) x = 0.5$ for any $m \in \Rn{3}$. It follows, by definition, that $A_s(m)$ cannot be negative semidefinite. In other words, $A_s(m)$ has at least one strictly positive eigenvalue for any coordinate shift $m \in \Rn{3}$.
    $\blacksquare$
\end{pf}

In summary, we showed that the three-dimensional system~\eqref{eq:counterexample} with energy-preserving quadratic nonlinearity has the following properties:
\begin{enumerate}
    \item The system is long-term bounded (Lemma~\ref{lemma:Counterexample_LongTermBoundedness}).
    \item The system has an effective nonlinearity (Lemma~\ref{lemma:Counterexample_EffectiveNonlinearity}).
    \item The matrix $A_s(m)$ has a positive eigenvalue for any coordinate shift $m \in \Rn{3}$ (Lemma~\ref{lemma:Counterexample_Eigenvalues}).
\end{enumerate}
This system provides a counterexample to Theorem~\ref{thm:necessity} proposed by Schlegel and Noack~\cite{schlegel2015long}. The system satisfies the premises of the theorem but does not satisfy the conclusion, as the symmetric linear part $A_s(m)$ has a positive eigenvalue for any coordinate shift $m$. Note that this implies that Theorem~\ref{thm:necessity} is false for $n\geq 3$ as higher dimensional counterexamples can be obtained by augmenting system~\eqref{eq:counterexample} with additional dynamics.


Lastly, we comment on the proof of Theorem~\ref{thm:necessity} provided in~\cite{schlegel2015long}. The proof relies expressing the system dynamics in a phase and magnitude form. The phase dynamics are then approximated for large state amplitudes by neglecting some terms. However, this original dynamics can be bounded while approximated dynamics fail to be bounded. This can be shown with the counterexample system~\eqref{eq:counterexample}. It is possible that the necessity theorem could be restated, with additional assumptions, to hold in higher dimensions. However, it is not clear how to make such an amendment and this is left for future work.

%% file: Content_Counter/Conclusions.tex
\section{Conclusion} \label{sec:conclusion}

This paper investigates the necessary condition proposed by Schlegel and Noack~\cite{schlegel2015long} for boundedness of quadratic systems with energy-preserving nonlinearity. Our main findings are twofold.
First, we verified that the necessary condition for boundedness holds for systems of this class with two dimensions.
Second, we constructed a three-dimensional system that is globally asymptotically stable and has effective nonlinearity, yet violates the necessary condition. This shows the condition does not generally hold in dimensions greater than two.
%
These results clarify the limitations and gaps of existing boundedness theory for quadratic systems with energy-preserving nonlinearities. 
Future work should focus on bridging this gap as many engineering applications rely on Theorem~\ref{thm:boundedness_sufficient}, which is only sufficient and can be overly conservative. 

%% file: Content_Counter/Appendix.tex
\appendix

\section{Proof of Lemma~\ref{lem:2D_rotation_invariance}} \label{sec:appendix_2DLemmaProof}

Recall the two-dimensional quadratic system in original coordinates~\eqref{eq:general_system}:
\begin{align}
    \frac{d}{dt} x(t) = c + L x(t) + (q^\top x(t)) \bmtx 0 & 1 \\ -1 & 0 \emtx x(t)
\end{align}
and the rotated system~\eqref{eq:rotated2D}:
\begin{align}
    \frac{d}{dt} \hat{x}(t) = \hat{c} + \hat{L} \hat{x}(t) + (\hat{q}^\top \hat{x}(t)) \bmtx 0 & 1 \\ -1 & 0 \emtx \hat{x}(t)
\end{align}
where $\hat{c} = R c$, $\hat{L} = R L R^\top$, and $\hat{q} = R q$ for some rotation matrix $R$. 

The trapping region SDP~\eqref{SDP:TR_sufficient} for the original system is given by:
\begin{align} 
    a^* = \min_{m \in \Rn{2}, a \in \Rn{}} a \quad \text{s.t.} \quad \frac{1}{2}(L+L^\top) - \sum_{i=1}^2 m_i Q^{(i)} \preceq a I,
\end{align}
where
\begin{align*}
    Q^{(1)} &= \frac{1}{2}\left( q \bmtx 0 & 1 \emtx + \bmtx 0 \\ 1 \emtx q^\top\right)
    &&= \begin{bmatrix}
        0 & 0.5q_1 \\
        0.5q_1 & q_2
    \end{bmatrix},
    \\
    Q^{(2)} &= \frac{1}{2}\left(q \bmtx -1 & 0 \emtx + \bmtx -1 \\ 0 \emtx q^\top\right)
    &&= \begin{bmatrix}
        -q_1 & -0.5q_2 \\
        -0.5q_2 & 0
    \end{bmatrix}.
\end{align*}
Negative semidefiniteness is invariant under congruence transformations. Thus, the feasiblity of the LMI is unchanged if we multiply on the left by $R$ and on the right by $R^\top$. This yields the following equivalent form for the LMI:
\begin{align}
 \frac{1}{2}(\hat{L}+\hat{L}^\top) - R\left( \sum_{i=1}^2 m_i Q^{(i)} \right) R^\top \preceq a I
\end{align}
This step uses $RR^\top=I$ and the definition of $\hat{L}$.

Next, note that the second term on the left-hand side of the LMI can be expressed as:
\begin{align*}
    R\left( \sum_{i=1}^2 m_i Q^{(i)} \right) R^\top 
    & = \frac{1}{2} R \left( q \bmtx -m_2 & m_1 \emtx + \bmtx -m_2 \\ m_1 \emtx q^\top \right) R^\top \\
    & = \frac{1}{2} \left( \hat{q} \bmtx -\hat{m}_2 & \hat{m}_1 \emtx + \bmtx -\hat{m}_2 \\ \hat{m}_1 \emtx \hat{q}^\top \right) 
\end{align*} 
where $\hat{m} := \bsmtx \hat{m}_1 \\ \hat{m}_2 \esmtx$ is defined so that $\bsmtx -\hat{m}_2 \\ \hat{m}_1 \esmtx := R \bsmtx -m_2 \\ m_1 \esmtx$.
This yields the following equivalent form for the original SDP:
\begin{align} 
    \hat{a}^* = \min_{\hat{m} \in \Rn{n}, \hat{a} \in \Rn{}} \hat{a} \quad \text{s.t.} \quad  \frac{1}{2}(\hat{L}+\hat{L}^\top) - \sum_{i=1}^2 \hat{m}_i \hat{Q}^{(i)}  \preceq \hat{a} I
\end{align}
This is exactly the form of the trapping region SDP for the transformed system.  Therefore, any feasible solution $(a,m)$ for the SDP with the original system corresponds to a feasible solution $(\hat{a}=a,\hat{m})$ for the SDP with the rotated system, and vice versa. This shows that the two SDPs are equivalent and have the same optimal cost, i.e., $a^* = \hat{a}^*$.